\documentclass[superscriptaddress, secnumarabic, amssymb, nobibnotes, aps, prl, reprint]{revtex4-2}

\usepackage{chemformula}
\usepackage{chemmacros}
\usepackage{graphicx}
\usepackage[breaklinks, colorlinks, citecolor = blue, linkcolor = black, urlcolor = black, pdftitle = {Anomalously Large Screening Lengths in Highly Concentrated Electrolytes: A Matter of Experimental Conditions}]{hyperref}
\usepackage[capitalise]{cleveref}
\usepackage[separate-uncertainty]{siunitx}
\usepackage{upgreek}
\usepackage{xspace}

\newcommand{\ie}{{\itshape i.e.,}\xspace}

\begin{document}

\title{Anomalously Large Screening Lengths in Highly Concentrated Electrolytes: \\ A Matter of Experimental Conditions}

\author{Thomas Tilger}
\author{Esther Ohnesorge}
\author{Michalis Tsintsaris}
\affiliation{Soft Matter at Interfaces, Department of Physics, Technical University of Darmstadt, 64289 Darmstadt, Germany}
\author{Nithya~Hellar}
\author{Kazue~Kurihara}
\affiliation{New Industry Creation Hatchery Center, Tohoku University, 980-8577 Sendai, Japan}
\author{Hayden Robertson}
\email{hayden.robertson@pkm.tu-darmstadt.de}
\author{Regine von Klitzing}
\email{klitzing@smi.tu-darmstadt.de}
\affiliation{Soft Matter at Interfaces, Department of Physics, Technical University of Darmstadt, 64289 Darmstadt, Germany}
\date{\today}

\begin{abstract}
Considerable debate exists about anomalous underscreening. 
Whilst surface force apparatus (SFA) measurements have confirmed anomalously large screening lengths ($\lambda$) in concentrated aqueous electrolytes, colloidal probe atomic force microscopy (CP-AFM) measurements have not. 
We demonstrate that by tuning experimental conditions, similarly large $\lambda$ can be achieved in concentrated aqueous \ch{LaCl3} electrolytes with CP-AFM.
Structural forces were also observed at high \ch{LaCl3} concentrations, which we suggest stem from the formation of diffuse ion clusters, and that the remaining counter-ions yield the large deduced $\lambda$.
\end{abstract}

\maketitle

\textit{Introduction} --- 
Electrostatic interactions and the arising electric double layer (EDL) force are the cornerstone of our understanding of interactions in colloidal systems \cite{Ninham.1999, Israelachvili.2011}.
According to the classical Debye-Hückel (DH) theory \cite{PeterDebye.1923}, dilute electrolytes can be described by an exponentially decaying force profile with the Debye length

\begin{equation}
    \lambda_{\mathrm{D}} = \sqrt{\frac{\varepsilon_{\mathrm{r}} \varepsilon_0 k_{\mathrm{B}} T}{2 e^2 N_{\mathrm{A}} I}}, \label{eqn:Debye_length}
\end{equation}

\noindent where $\varepsilon_{\mathrm{r}} \varepsilon_0$ represents the permittivity of the medium, $k_{\mathrm{B}}$ and $N_{\mathrm{A}}$ are the Boltzmann and Avogadro constants, respectively, $T$ is the absolute temperature, and $e$ is the elementary charge.
$I = \sum_i c_i z_i^2$ denotes the ionic strength of the electrolyte with respective valences $z_i$ and concentrations $c_i$.
\cref{eqn:Debye_length} demonstrates that $\lambda_\mathrm{D}$ decreases with increasing ionic strength.
In highly concentrated electrolytes, however, the underlying mean-field assumption in the DH theory of non-interacting ions breaks down.
As initially emphasized by Kirkwood \cite{Kirkwood.1936,Kirkwood.1954}, excluded volume effects must be accounted for at higher electrolyte concentrations, which ultimately cause the electrostatic interaction to transition from a monotonic exponential to a damped oscillatory decay \cite{Elliott.2024}.
In this higher concentration regime,
the experimentally observed screening length $\lambda$ increases with concentration, deviating from DH theory.

Recent studies have highlighted the ubiquitous nature of this \emph{underscreening effect}, which commonly manifests as re-entrant behavior with increasing electrolyte concentration \cite{Yuan.2022, Dunlop.2026, Dunlop2026b}.
The explanation for long-ranged interactions in highly concentrated electrolytes, however, is still under heavy debate and challenges our fundamental understanding of electrolytes \cite{Zeman.2021,Elliott.2024,Zhang.2024,Cross.2026}.
Motivated by a seminal analysis by Lee \emph{et al.} \cite{Lee.2017}, $\lambda$ derived from experiment and theory are often categorized according to a scaling law of type 

\begin{equation}
    \frac{\lambda}{\lambda_{\mathrm{D}}} \sim \left(\frac{a}{\lambda_{\mathrm{D}}}\right)^p, \label{eqn:Scaling_law}
\end{equation}

\noindent with ion size $a$ and exponent $p$.
\emph{Normal} underscreening, as deduced from the re-entrant swelling behavior of polymer brushes \cite{Robertson.2023}, small-angle X-ray scattering \cite{Dinpajooh.2024}, and optical second harmonic scattering \cite{Rondepierre.2025} is characterized by an exponent $1 \le p \le 2$.
In contrast, substantially larger $\lambda$ were deduced from surface force apparatus (SFA) measurements \cite{Pashley.1984c,Pashley.1984b,Baimpos.2014,Smith.2016}, and studies detecting the surface-excess of fluorescein \emph{via} fluorescence emission \cite{Gaddam.2019} across a variety of highly concentrated aqueous salt solutions.
Comparable findings in ionic liquids \cite{Gebbie.2013,Gebbie.2015,Smith.2016,Fung.2023} indicate that these anomalously large $\lambda$, coined \emph{anomalous} underscreening, are characterized by the exponent $p = 3$ in \cref{eqn:Scaling_law}, and represent a general feature of strongly correlated electrolytes under confinement.
Despite the conceptual similarities between the colloidal probe atomic force microscope (CP-AFM) and SFA, comparable long-ranged repulsive forces in highly concentrated aqueous electrolytes have not been found by AFM \cite{Kumar.2022}. 

Scaling theories have attributed anomalously large $\lambda$ in concentrated electrolytes to clusters of correlated ions \cite{Safran.2023}.
Most recently, polarizable molecular dynamics simulations by Baker~\textit{et~al.} extended preceding statistical mechanical analyses to reveal \emph{multiple} decay modes of the electrostatic potential, each of which with its own decay (screening) length and some closely matching the experimental results \cite{Baker.2026}.
That is, the authors prescribe that both \emph{normal} and \emph{anomalous} underscreening, along with further modes, are always present but can only be revealed by certain experimental approaches.
As such, we use the terms \emph{normal} and \emph{anomalous} strictly as operational labels, independent of their conventional meanings, to denote screening as governed by scaling with $p = \numrange{1}{2}$ and $p = 3$, respectively.

In the present work, we systematically converge the experimental conditions of CP-AFM towards those of an SFA to demonstrate that the experimental conditions strongly influence the interactions encountered in highly concentrated electrolytes.
We first revisit a conventional AFM setup, and then vary the size of the colloidal probe, equilibration time, and approach speed to reveal long-ranged repulsion in highly concentrated lanthanum chloride (\ch{LaCl3}) electrolytes.
Finally, we reveal the presence of structural forces at high \ch{LaCl3} concentrations, suggesting that the long-range decay stems from the formation of diffuse ion clusters and follows the scaling relationship previously only observed in SFA experiments \cite{Lee.2017}.

\begin{figure}[tb!]
    \begin{center}
	\includegraphics[width=\columnwidth]{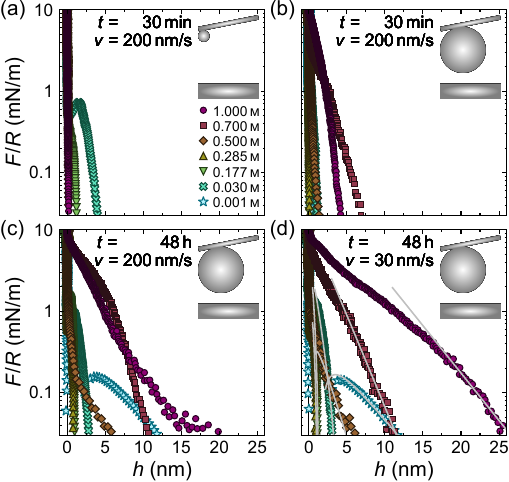}
	\caption{Normalized interaction force $F / R$ between a silica CP and a planar quartz substrate at separation $h$ immersed in aqueous \qtyrange{0.001}{1.000}{\Molar} \ch{LaCl3}.
    Top row: Data captured after an equilibration time of $t=\SI{30}{\minute}$ with an approach speed of $v=\SI{200}{\nano\metre/\second}$ using a CP of (a) $R=\SI{2.1}{\micro\meter}$ and (b) $R=\SI{8.5}{\micro\meter}$.
    Bottom row: Data captured with the larger CP $R=\SI{8.5}{\micro\meter}$ after a total equilibration time of $t=\SI{48}{\hour}$ and 
    approach speed of (c) $v=\SI{200}{\nano\metre/\second}$ and (d) $v=\SI{30}{\nano\metre/\second}$.
    The $F/R$ profiles present an emerging repulsion in the highly concentrated regime with (a)~$\rightarrow$~(b) increasing $R$, (b)~$\rightarrow$~(c) longer $t$, and (c)~$\rightarrow$~(d) decreasing $v$.
    The solid lines in (d) denote fitted DLVO models.
    A linear representation of this figure is provided as Fig.~S1 \cite{SuppMat}. \label{fig:Fig1}}
    \end{center}
\end{figure}

\textit{Results and Discussion} --- 
To commence, we employed the standard CP-AFM setup utilized by previous investigations of electrostatic interactions \cite{Valmacco.2016b, Smith.2019, Ludwig.2021b, Tilger.2026}: colloidal probe radius $R =\SI{2.1}{\micro\metre}$, equilibration time $t= \SI{30}{\minute}$, approach speed $v= \SI{200}{\nano\metre/\second}$.
The data in \cref{fig:Fig1}\,(a) presents the force curves obtained under these conditions, whereby the EDL repulsion is suppressed with increasing $c$.
However, for $c \ge \SI{0.500}{\Molar}$, pure attractive van der Waals forces are observed down to contact as EDL forces are completely screened; this is highlighted in Fig.~S1 \& Fig.~S2 \cite{SuppMat}.
\nocite{Valmacco.2016b, Lee.2017, Robertson.2023, Shannon.1976, RuizCabello.2014, Valmacco.2016, Ludwig.2021b, Hasted.1948, Wyman.1938, Malmberg.1956, Uematsu.1980}
These results align with the expectations from the canonical DH theory and a previous CP-AFM study \cite{Kumar.2022}.
This represents the current state of the art regarding the reported absence of anomalous underscreening in AFM measurements.

To converge CP-AFM towards SFA, we consider the inherent differences in measuring geometries; namely, the difference in radii of curvature $R$ between the interacting surfaces.
While it is technically infeasible to use CPs with dimensions comparable to the cm-scale SFA cylinders, we move towards a larger contact area by increasing $R$ from \SI{2.1}{\micro\meter} to \SI{8.5}{\micro\meter}.
Both $t$ and $v$ are kept constant at \SI{30}{\minute} and \SI{200}{\nano\metre/\second}, respectively, as to only interrogate the geometrical influence.
Moving from \cref{fig:Fig1}\,(a) to (b), a clear transition in the detected interaction force is observed for the higher $c$: from a pure attraction towards a repulsion over several nanometers.
These results clearly contradict the DH predictions, which state $\lambda_\mathrm{D} < \SI{0.2}{\nano\metre}$ for $c \ge \SI{0.500}{\Molar}$ electrolytes of \ch{LaCl3}.
Thus, true re-entrant behavior of repulsive interactions is observed with increasing $c$. 
Earlier studies suspected a lack of sensitivity to be responsible for the absence of long-ranged repulsive forces in AFM experiments \cite{Gebbie.2013,Baimpos.2014}.
The increase in colloidal probe size here indeed suppresses the noise floor, however, the magnitude of normalized electrostatic force encountered in SFA studies exhibiting an anomalously large electrostatic repulsion falls within the resolution of our AFM measurements \cite{Gebbie.2013,Baimpos.2014,Gebbie.2015,Smith.2016}.
Additionally, the repulsion encountered in \cref{fig:Fig1}\,(b) upon increasing $R$ would have been clearly resolvable with the small probe in \cref{fig:Fig1}\,(a).
Therefore, we conclude that the decisive factor presented here is indeed the radius of curvature of the interacting surfaces rather than the improved force resolution.

Having examined the geometric properties, we now turn to the impact of equilibration time $t$ on the resultant interaction forces.
Previous SFA experiments have shown the evolution of long-ranged interactions in diluted \ch{LaCl3} electrolytes over several hours due to slow diffusion processes \cite{Pashley.1984b}.
Complementary measurements on highly concentrated systems involving ionic liquids show that the electrolyte structure near interfaces evolves over comparable timescales \cite{Han.2020}, altering the extent of long-range interactions \cite{Fung.2023}.
Therefore, the preceding setup with the $R=\SI{8.5}{\micro\meter}$ was left to equilibrate for at least \SI{48}{\hour}. 
As can be seen in \cref{fig:Fig1}\,(c), the re-entrant behavior of repulsive interactions is increased. 
Thus, the changes in the interaction force over two days could indicate very slow ion dynamics near the interfaces; this aligns with previous findings \cite{Fung.2023}.
We hypothesize that the strong ion-ion correlations in these highly concentrated \ch{LaCl3} solutions could slow down diffusive processes close to the interfaces, resulting in the requisite longer equilibration times. 
Importantly, similarly slow processes could occur during the formation of hydrated ion clusters \cite{Safran.2023}.  

Finally, we illuminate the role of approach speed $v$.
While the EDL force is considered a static equilibrium force, any object moving through a fluid is subject to dynamic viscous forces.
Due to the larger geometrical dimensions which inevitably enhance hydrodynamic effects, the mica cylinders in SFA experiments are typically moved at much slower speeds than CPs: $\sim \SI{1}{\nano\metre/\second}$ \textit{vs.} $\sim \SI{200}{\nano\metre/\second}$.
To mimic these conditions, $v$ was reduced to \SI{30}{\nano\metre/\second}; a further reduction in $v$ was limited by the vibration isolation of the AFM. 
$R$ and $t$ remained unchanged.
As depicted in \cref{fig:Fig1}\,(d), the interaction is clearly altered by the reduction in $v$.
Importantly, we stress that the impact of hydrodynamic forces is negligible on the system investigated; 
see Appendix~B for further information.
Small changes are observed at low to intermediate $c$, while the main impact is exerted on the force curves corresponding to the highly concentrated \ch{LaCl3} solutions.
The overall trend of re-entrant double layer repulsion is, however, reinforced.

To this point, we have systematically converged the experimental conditions of our CP-AFM setup toward those of an SFA, demonstrating the crucial impact of measurement conditions on the interactions observed in highly concentrated electrolyte solutions.
Specifically, by employing a large $R$, long $t$, and slow $v$, CP-AFM is capable of detecting long-ranged repulsive forces similar to the SFA.
These results align with simulations from Baker et al., who suggest the omnipresence of multiple decay (screening) lengths \cite{Baker.2026}.
However, we emphasize that none of the aforementioned parameters alone are responsible for the emergent repulsion at high $c$; rather, the observed long-ranged repulsive forces are the result of the synergy of all parameters.
This is further exemplified in Fig.~S3, which presents the absence of repulsive interactions for the small $R$ under the same conditions 
as in \cref{fig:Fig1}\,(d).

To quantify the resultant screening lengths $\lambda$, the theory of Derjaguin, Landau, Verwey, and Overbeek (DLVO) was employed \cite{Derjaguin.1939,DerjaguinB.V.L.Landau.1941,E.J.W.Verwey.1948} as a superposition of the 
EDL repulsion according to the DH approximation \cite{Butt.2018}
\begin{equation}
    \frac{ F_{\mathrm{EDL}} }{ R } = 4 \pi \varepsilon_{\mathrm{r}} \varepsilon_0 \frac{ \psi_{ \mathrm{eff}}^2 }{ \lambda } e^{ - \frac{ h }{ \lambda } } \label{eqn:EDL_Force}
\end{equation}
and the non-retarded van der Waals attraction \cite{Hamaker.1937}
\begin{equation}
    \frac{ F_{\mathrm{vdW}} }{ R } = - \frac{H}{6 h^2}. \label{eqn:vdW_Force}
\end{equation}
In line with literature, the outermost region of the force curves was modeled \cite{MontesRuizCabello.2015, Trefalt.2016} and the Hamaker constant $H = \SI{2.25(50)e-21}{\joule}$ was deduced \cite{Valmacco.2016}. 
Further information on the DLVO modeling is seen in \S S3 \& \S S7.

\begin{figure}[tb!]
    \begin{center}
	\includegraphics[width=\columnwidth]{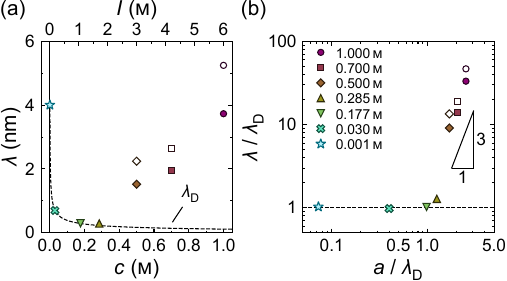}%
	\caption{The experimental screening length $\lambda$ of aqueous \ch{LaCl3} electrolytes as extracted from model fits of the force curves in \cref{fig:Fig1}\,(d) (closed symbols). 
    (a) $\lambda$ as a function of concentration $c$ and ionic strength $I = 6 c$. The dashed black line presents the predictions of the DH theory.  
    (b) $\lambda$ normalized by the Debye length $\lambda_{\mathrm{D}}$ as a function of the normalized mean ion size $a / \lambda_{\mathrm{D}}$. 
    Open symbols: From fits taking the oscillatory contribution into account (see \cref{fig:Fig3}). 
    A linear representation of the data and model fit of the scaling law from \cref{eqn:Scaling_law} are presented in Fig.~S6 \cite{SuppMat}. \label{fig:Fig2}}
    \end{center}
\end{figure}

\Cref{fig:Fig2}\,(a) presents the experimentally determined screening length $\lambda$ from the force curves in \cref{fig:Fig1}\,(d) as a function of \ch{LaCl3} concentration, and mirrors the non-monotonic behavior observed across the full concentration range. 
For $c \le \SI{0.177}{\Molar}$, the DLVO model captures the data well, and $\lambda$ decreases monotonically in agreement with $\lambda_{\mathrm{D}}$.
This validates the modeling approach and identifies the observed repulsion as the EDL force.
However, as the electrolyte concentration increases further, systematic deviations between the model and the measured force curves in \cref{fig:Fig1}\,(d) become apparent, indicating that the repulsion cannot be simply explained by a reemergence of the EDL force.
Nevertheless, to ensure a consistent analysis across all concentrations and facilitate comparison with previous studies, the same model was applied at higher $c$.
For $c \ge \SI{0.285}{\Molar}$, a monotonic increase in $\lambda$ is observed, exceeding $\lambda_{\mathrm{D}}$ by several \si{\nano\metre} and also surpassing the obtained $\lambda$ in the \si{\milli\Molar} range. 
Additionally, the magnitude of the deduced effective surface potential $\psi_{ \mathrm{eff}}$ exhibits a turning point at $c \approx \SI{0.285}{\Molar}$ (see Fig.~S7), which may occur due to the adsorption of $\ch{La}^{3+}$ at the negatively charged confining surfaces.

\begin{figure*}[tb!]
    \begin{center}
    \includegraphics[width=0.65\textwidth]{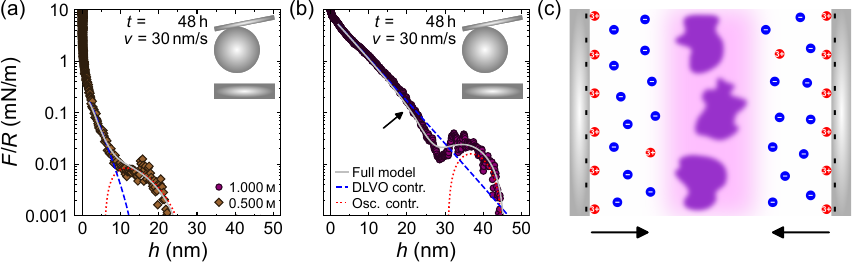}
	\caption{Normalized interaction force $F / R$ between the \SI{8.5}{\micro\meter} silica probe and quartz substrate in (a) \SI{0.500}{\Molar} and (b) \SI{1.000}{\Molar} \ch{LaCl3}. 
    Data are recast from \cref{fig:Fig1}\,(d) over an extended force range. 
    Data corresponding to \SI{0.700}{\Molar} \ch{LaCl3} is presented in Fig.~S8 \cite{SuppMat}. 
    The arrow in (b) highlights the deviation from linearity; i.e., between the DLVO contribution and the employed extended model.
    (c) Visual representation of the hypothesized structure in the slitpore under confinement in the high concentration regime: (red) \ch{La^{3+}} cations adsorbed at the (negative) quartz substrate (inducing charge reversal; see Fig.~S7) and (blue) \ch{Cl-} anions governing the EDL contribution; ordered (purple) ion clouds in a central diffuse zone inducing the structural force.
    \label{fig:Fig3}}
    \end{center}
\end{figure*}

Next, we place our observations in a broader context through a scaling analysis of the extracted $\lambda$.
SFA measurements have consistently demonstrated that across aqueous solutions of various alkali halides and also ionic liquids, $\lambda$ can be normalized to collapse onto a single curve \cite{Smith.2016,Lee.2017,Fung.2023} as described by the scaling relationship expressed in \cref{eqn:Scaling_law} with $p = 3$ \cite{Lee.2017}.
This is in contrast to other investigations which align with $p \le 2$ \cite{Robertson.2023, Dinpajooh.2024, Rondepierre.2025}.
Here we apply this scaling law to the deduced $\lambda$ presented in \cref{fig:Fig2}\,(a), normalizing $\lambda$ to $\lambda_\mathrm{D}$ (see \cref{eqn:Debye_length}).
As illustrated by the slope of the black triangle in \cref{fig:Fig2}\,(b), the data presented in this work are indeed consistent with the proposed scaling law for \emph{anomalous} underscreening from \cref{eqn:Scaling_law} with $p=3$.
A visualization of the smaller scaling exponents $p = \numrange{1}{2}$, as expected for \emph{normal} underscreening, is presented in Fig.~S6. 
The onset of the increase in $\lambda/\lambda_\mathrm{D}$ coincides with $\lambda_{\mathrm{D}} \approx a$, highlighting the remarkable similarity between the AFM results presented herein and prior SFA investigations. 
Of note, this similarity differentiates the current work from previous AFM investigations \cite{Kumar.2022} that could not confirm the long-ranged repulsion observed in SFA experiments.

The results presented naturally raise the question of the origin of the long-ranged repulsion at high concentrations.
As noted, the DLVO model fails to fully capture the measured forces at high $c$ due to the pronounced curvature (i.e., the deviation from linearity as in \cref{fig:Fig1}\@(d)). 
As such, select data from \cref{fig:Fig1}\@(d) are recast over an extended force range in \cref{fig:Fig3}, revealing an oscillation at low forces.
The characteristic shape of these force curves motivates the inclusion of an exponentially damped oscillatory contribution with a depletion of the form

\vspace{-1em}
\begin{equation}
    \small
	\frac{ F_{\mathrm{struc}} }{ R } =
	\begin{cases}
		F_{\mathrm{depl}} / R  = F_\mathrm{osc} (h^*) / R & h < h^{\ast} \\
		F_{\mathrm{osc}} / R = A e^{ - \nicefrac{ h }{ \xi } } \cos\left(  {2\pi h/ }{ \Lambda } + \theta \right) & h \ge h^{\ast}
	\end{cases} \label{eqn:OSC_Force}
\end{equation}

\noindent with amplitude $A$, correlation length $\xi$, oscillation wavelength $\Lambda$ and phase shift $\theta$ \cite{Klapp.2008, MoazzamiGudarzi.2016, Ludwig.2021b, Ludwig.2022}.
$h^{\ast}$ denotes the separation at which the $F_\mathrm{osc}$ transitions continuously into a $F_\mathrm{depl}$ attraction \cite{Ludwig.2021b}.
The total structural force can then be described as F$_\mathrm{struc}$ = F$_\mathrm{depl}$ + F$_\mathrm{osc}$; see \S S6 in the SM for further details \cite{SuppMat}.
Combined with the DLVO contribution, this extended model (\cref{fig:Fig3}) describes the experimental data well over the entire force range \cite{Ludwig2021}. 
The corresponding screening lengths from the EDL repulsion were extracted from the fits of the extended model, and are presented as open symbols in \cref{fig:Fig2}.

In addition to the previous question regarding the origin of the large $\lambda$, this new representation prompts an additional question: what is the origin of the oscillation?
We hypothesize that the two are intrinsically linked.
To interpret these results, we draw inspiration from previous works which describe the energetically favorable formation of ion clusters at high $c$ \cite{Safran.2023, Hartel.2023}.
In both works, ion clusters are ascribed to possess ``complex topologies'' with either a neutral or negligible charge.
In the present work, we define ion clusters as \emph{mechanically fragile clouds of correlated ions}, and separate the discussion into structural and electrostatic contributions; an artistic interpretation of the system is presented in \cref{fig:Fig3}\,(c).

\textit{Structural contribution}: 
We suggest that the origin of the oscillation observed in \cref{fig:Fig3} stems from the expulsion of diffuse clouds of clustered cations and anions from the slitpore during confinement. 
First, a weak repulsion occurs when the ion clusters are slightly compressed.
Upon further compression to $h^*$, all clusters are then ``expelled'', resulting in attractive depletion force; however, we note that the net force in this region is dominated by the electrostatic repulsion between the confining surfaces. 
This interplay results in depletion forces which partially counteract the electrostatic repulsion, yielding in the curvature as observed in \cref{fig:Fig3} (and \cref{fig:Fig1}\,(d)); this is highlighted by the arrow in \cref{fig:Fig3}\,(b).
In contrast to former studies on well-ordered nanoparticles \cite{Ludwig2021}, only one oscillation is observed here for solutions of \SIlist{0.500;1.000}{\Molar} \ch{LaCl3}.
The absence of additional observable oscillations could originate from the intrinsic low degree of ordering amongst the ion clusters (i.e., size polydispersity) confined between the two surfaces under each approach as well as the low charge of the ion clusters. 
This correlation between ordering under confinement and particle charge has previously been shown by \citeauthor{Ludwig.2022} \cite{Ludwig.2022}. 
Interestingly, with increasing $c$, 
the oscillation becomes more pronounced, which indicates a less diffuse zone of ion clusters \cite{Safran.2023}.

\textit{Electrostatic contribution}:
The modeling of the surface potential indicates a charge reversal (Fig.~S6), which is induced by the adsorption of the positively charged \ch{La^{3+}} ions at the negatively charged surfaces.
As such, we assume that the majority of free ions are \ch{Cl-} (i.e., the counter-ions), and thus not all ions are involved in cluster formation; see \cref{fig:Fig3}\,(c). 
It is these free ions that contribute to the electrostatic screening and govern the observed long-range $\lambda$.
As $\lambda$ is large, we assume that the $c$ of the free ions (i.e., $c_\mathrm{free}$) is quite small.
Counterintuitively, $\lambda$ increases with increasing $c$, such that $c_\mathrm{free}$ decreases; that is, there are more free ions for \SI{0.500}{\Molar} \ch{LaCl3} than for \SI{1.000}{\Molar} \ch{LaCl3}. 
Under the assumption that $\lambda = \lambda_\mathrm{D}$, we calculate the ionic strength of the free ions as \qtylist{0.017; 0.012; 0.003}{\Molar} for \ch{LaCl3} solutions of \qtylist{3.0; 4.2; 6.0}{\Molar} ionic strength, respectively (see Appendix~C for further information).
Building on this idea and uniting the structural and electrostatic contributions, we conclude that the observed long-range repulsion is directly linked to the structure formed by the ion clusters, such that increasing $\lambda$ with increasing $c$ is directly correlated with an enhanced formation of clusters, which could be due to the rapidly reducing solvent availability with increasing $c$. 

\textit{Conclusion} --- 
We have demonstrated through colloidal-probe atomic force microscopy measurements (CP-AFM) that interactions in highly concentrated aqueous \ch{LaCl3} electrolytes are not governed exclusively by ionic strength.
By converging AFM experimental conditions towards those of the SFA, we have systematically tuned the interaction from purely attractive van der Waals forces to long-ranged repulsive interactions.
In this context, the size of the interacting surfaces and the equilibration time were increased, and the approach speed was decreased.
Notably, at higher electrolyte concentrations the repulsion increased and the extracted screening lengths followed the universal scaling law previously inferred from SFA experiments \cite{Lee.2017}.
Thus, our findings mark the first experimental observation of anomalous underscreening (\ie $p=3$) in aqueous electrolytes with the AFM, challenging former beliefs that anomalously large screening lengths are \emph{inherently} undetectable by AFM. 

Furthermore, weak structural forces were observed at high electrolyte concentrations, which we interpret as the formation of fragile clouds of correlated ions.
For these concentrations, DLVO theory failed to accurately capture the experimental data, and the model was extended to include a structural contribution.
We suggest that the observed long-range repulsion originates from the remaining free ions post cluster formation.
The formation of these structures is slow, taking 2 days in the present study. 
Due to the diffuse nature of this cluster layer, a slow approach speed is required to reveal their ordering.
As such clustering was not detectable by bulk scattering experiments \cite{Elliott.2024}, we hypothesize that these large structures are induced by surfaces. 
Our observations point to an important role of electrolyte structuring in the emergence of long-ranged repulsive interactions at high electrolyte concentrations. 
Specifically, the formation of such structuring requires a certain size of confinement as well as long equilibration 
times.
This work highlights avenues for systematically controlling the strength of the repulsive interactions in highly concentrated aqueous electrolytes, with direct implications for the stability of biological systems under extreme conditions and for the design of next-generation high-concentration battery electrolytes.

\begin{acknowledgments}
\textit{Acknowledgments} --- H. R. gratefully acknowledges a Humboldt Research Fellowship from the Alexander von Humboldt Stiftung.
In no particular order, the authors would like to thank Christian Holm, Patrick Kekicheff, Frieder Mugele, Alexander Smith, and Erica Wanless for fruitful discussions.

\end{acknowledgments}

\textit{Data availability} ---
Supporting data and analysis scripts will be publicly available after embargo.

\bibliography{bibliography}

\appendix*
\section{End matter}

\textit{Appendix A: Experimental details ---}
Colloidal probe atomic force microscopy (CP-AFM) measurements \cite{Butt.1991,Ducker.1991} were conducted between a silica microsphere attached to a cantilever and a planar quartz substrate (Plano) in aqueous \ch{LaCl3} (Sigma-Aldrich, \SI{99.999}{\percent}) solutions over a wide range of concentrations (\qtyrange{0.001}{1}{\Molar}).
Ultrapure Milli-Q water (Merck, \SI{18.2}{\mega\ohm\centi\metre} resistivity at \SI{25}{\celsius}) was used throughout unless stated otherwise.
The quartz disk was initially cleaned in a \num{3}:\num{1} (vol:vol) piranha solution of sulfuric acid (Carl Roth, \SI{98}{\percent}) and hydrogen peroxide (Carl Roth, \SI{30}{\percent}) for \SI{30}{\minute} to remove any organic contaminants.
The tipless cantilever (CSC37, MikroMasch, Estonia) was initially cleaned with water and ethanol (Carl Roth, \SI{99.9}{\percent}) followed by an oxygen plasma treatment (\SI{0.4}{\milli\bar}, \SI{50}{\watt}) for \SI{210}{\second}.
To prepare the colloidal probes, a silica microsphere was glued to the end of the cantilever with UHU PLUS ENDFEST adhesive using a micromanipulator (Sutter Instrument).
The colloidal probe was then sintered at \SI{1150}{\celsius} for \SI{3}{\hour}; this process removes organic contaminants which may remain from the glue and reduces the surface roughness of the microspheres \cite{Valmacco.2016}.
Two sizes of nonporous silica microspheres were employed; their post-sintering radii were determined from scanning electron microscopy images to be \SI{2.1\pm0.1}{\micro\metre} (Bangs Laboratories) and \SI{8.5\pm0.1}{\micro\meter} (microParticles).
The spring constant of each individual cantilever / colloidal probe ($k \approx \SI{0.8}{\newton / \metre}$) was extracted from its thermal noise spectrum via the Sader method \cite{Sader.1999}.

Prior to each measurement, the quartz disk and colloidal probe were cleaned with water and ethanol, and again treated with an oxygen plasma for \SI{210}{\second}.
For all experiments, a temperature of \SI{30}{\celsius} was maintained within the sample chamber of the closed-loop AFM (MFP-3D, Oxford Instruments), and the chamber was fully sealed in order to prevent evaporation of the electrolyte during the experiments.
This is confirmed by the close agreement between theoretical Debye lengths $\lambda_\mathrm{D}$ and the experimentally observed decay lengths $\lambda$ for low electrolyte concentrations.
A setpoint of \SI{120}{\nano\newton} was employed for all measurements, which is justified in Fig.~S4.
The travel distance of the $z$-piezo was set to \SI{600}{\nano\meter}.
Under the experimental conditions within this work, silica and quartz bear a negative charge due to ionization of surface silanol groups (\ch{SiO-}) \cite{Iler.1979,Kosmulski.2023}.
However, strong adsorption of the trivalent \ch{La^{3+}} ions is expected to induce charge reversal in the investigated concentration range \cite{Besteman.2004}.
The reported interaction forces are normalized by the colloidal probe radius $R$ according to Derjaguin's approximation $F / R \left( h \right) = 2 \pi W ( h )$ to obtain the interaction energy $W$ of two planar surfaces at the same separation $h$ \cite{Derjaguin.1934}.

\textit{Appendix B: Role of solution viscosity ---}
It is important to consider the role of hydrodynamic forces and viscosity in these solutions; in particular the impact of approach speed and bead size.
So we pose the question: to which extent is electrolyte viscosity responsible for the velocity dependence observed in our experiments?
The role of viscous forces in SFA measurements has been examined in detail, with results demonstrating that the ratio between electrostatic and viscous forces scales as $\eta R v$, with dynamic viscosity $\eta$, radius of curvature $R$, and approach speed $v$ \cite{Lhermerout.2018}.
Under the experimental conditions employed in this work, previous investigations on the \ch{LaCl3} electrolyte have reported a monotonic increase in viscosity with increasing concentration, with a maximum of approximately \SI{1.6}{\milli\pascal\second} at \SI{1}{\Molar} \cite{Isono.1984, Liu.2009}.
Importantly, this is not more than twice the viscosity of pure water.
Furthermore, compared to our AFM investigations with a $R =  \SI{8.5}{\micro\metre}$ colloidal probe approaching the surface at $v \le \SI{200}{\nano\metre/\second}$, the influence of viscosity on SFA experiments ($\eta = \SI{157}{\milli\pascal\second}$, $R = \SI{3.3}{\milli\metre}$, $v \ge \SI{0.015}{\nano\metre/\second}$ in \cite{Cross.2026} and $\eta = \SI{37}{\milli\pascal\second}$, $R = \SI{8.9}{\milli\metre}$, $v \ge \SI{0.07}{\nano\metre/\second}$ in \cite{Lhermerout.2018}) is at least \num{2} and \num{8} times as larger, respectively.
Hydrodynamic forces arising from fluid drainage can therefore be neglected for the experiments presented here at \SI{200}{\nano\metre/\second}.
The changes observed upon reducing the approach speed must originate from mechanisms beyond simple viscous drag, which would instead predict stronger repulsion at higher approach speeds.
We propose that the emergence of long-ranged repulsive interactions requires the presence of an equilibrated ionic structure in the vicinity of the interfaces, which is perturbed if the interfaces approach too fast.
Observations of ordered structure formation in confined ionic liquids support this interpretation \cite{Ueno.2010, Canova.2014, FedericiCanova.2015, Tomita.2018}.
We propose that the emergence of long-ranged repulsive interactions requires the presence of an equilibrated ionic structure in the vicinity of the interfaces, which is perturbed if the interfaces approach too fast.

\textit{Appendix C: Calculation of free-ion ionic strength}
As stated in the main text, we hypothesize that the long-range electrostatic repulsion originates from the low concentration of the free-ions remaining after clustering.
We suggest that the majority of the free ions are the \ch{Cl-} counter-ions.
Under the assumption that $\lambda = \lambda_\mathrm{D}$,
we calculate the ionic strength of the remaining free ions according to \cref{eqn:Debye_length}.
These results are presented in Table~\ref{freeions}.

\begin{table}[htb!]
    \centering
    \begin{tabular}{c|c|c}
        $c$ (M) & $I$ (M) & $I_\mathrm{free}$ (M) \\ \hline
        0.5 & 3.0 & 0.017 \\
        0.7 & 4.2 & 0.012 \\
        1.0 & 6.0 & 0.003
    \end{tabular}
    \caption{Summary of concentrations $c$ and ionic strengths $I$ for select \ch{LaCl3} solutions. 
    The ionic strength of the free ions remaining after clustering $I_\mathrm{free}$ is also presented.
    }
    \label{freeions}
\end{table}

\end{document}


\title{Supplemental Material for \\ ``Anomalously Large Screening Lengths in Highly Concentrated Electrolytes: A Matter of Experimental Conditions''}

\author{Thomas Tilger}
\author{Esther Ohnesorge}
\author{Michalis Tsintsaris}
\affiliation{Soft Matter at Interfaces, Department of Physics, Technical University of Darmstadt, 64289 Darmstadt, Germany}
\author{Nithya~Hellar}
\author{Kazue~Kurihara}
\affiliation{New Industry Creation Hatchery Center, Tohoku University, 980-8577 Sendai, Japan}
\author{Hayden Robertson}
\email{hayden.robertson@pkm.tu-darmstadt.de}
\author{Regine von Klitzing}
\email{klitzing@smi.tu-darmstadt.de}
\affiliation{Soft Matter at Interfaces, Department of Physics, Technical University of Darmstadt, 64289 Darmstadt, Germany}
\date{\today}

\maketitle

\clearpage

\section{Additional representations of the force spectroscopy data}

\begin{figure}[htb!]
    \begin{center}
    \includegraphics[width=5.0625in]{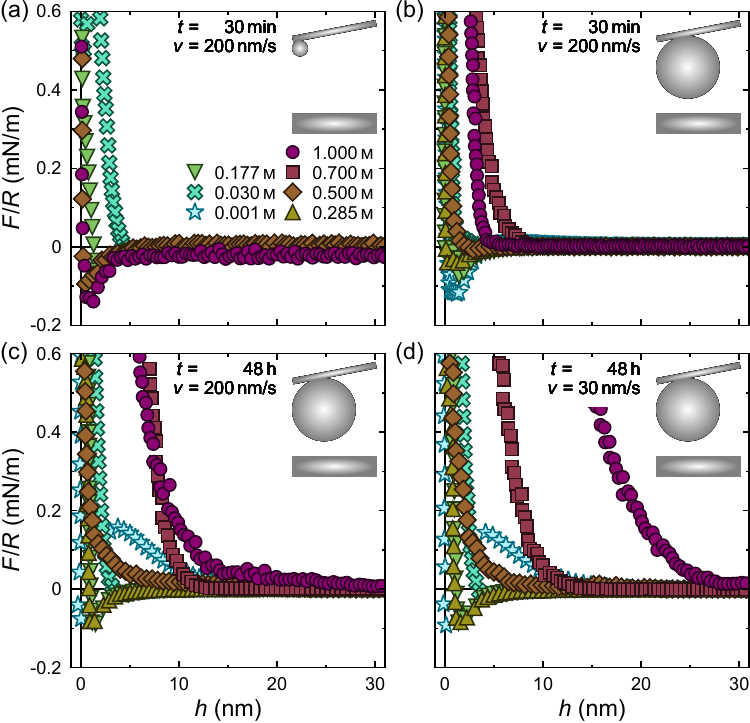}
    \caption{Normalized interaction force $F / R$ between a silica colloidal probe and a planar quartz substrate at separation $h$ immersed in aqueous \ch{LaCl3} electrolytes of various concentrations: \SIlist{0.001; 0.030; 0.177; 0.285; 0.500; 0.700; 1.000}{\Molar}.
    Top row: Data captured after an equilibration time of \SI{30}{\minute} with an approach speed of \SI{200}{\nano\metre/\second} using a (a) \SI{2.1}{\micro\meter} and (b) \SI{8.5}{\micro\meter} colloidal probe.
    Bottom row: Captured with the larger \SI{8.5}{\micro\meter} colloidal probe after an equilibration time of \SI{48}{\hour} at an approach speed of (a) \SI{200}{\nano\metre/\second} and (b) \SI{30}{\nano\metre/\second}.
    This figure is a linear representation of Fig.~1 in the main text and visualizes attractive and repulsive interactions. \label{fig:Suppmat_Fig1_Linear}}
    \end{center}
\end{figure}

In this section we present additional representations of the force spectroscopy data shown in the main text.
\cref{fig:Suppmat_Fig1_Linear} displays a linear representation of Fig.~1 to highlight the attractive interactions in aqueous \ch{LaCl3} solutions that are obscured on a logarithmic $y$-axis.

\clearpage

\begin{figure}[htb!]
    \begin{center}
    \includegraphics[width=5.0625in]{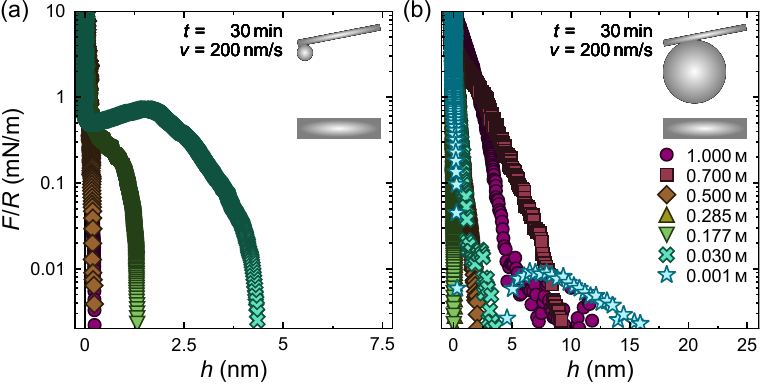}
    \caption{Normalized interaction force $F / R$ between a silica colloidal probe and a planar quartz substrate at separation $h$ immersed in aqueous \ch{LaCl3} electrolytes of various concentrations: \SIlist{0.001; 0.030; 0.177; 0.285; 0.500; 0.700; 1.000}{\Molar}.
    The data were captured after an equilibration time of \SI{30}{\minute} with an approach speed of \SI{200}{\nano\metre/\second} using a (a) \SI{2.1}{\micro\meter} and (b) \SI{8.5}{\micro\meter} colloidal probe.
    This figure is identical to Fig.~1\,(a) and (b) in the main text, but provides modified axes ranges. Only \SIlist{0.030; 0.177; 0.500; 1.000}{\Molar} are presented in (a).}
    \label{fig:Suppmat_Fig1_Alternative}
    \end{center}
\end{figure}

\cref{fig:Suppmat_Fig1_Alternative}\,(a) presents a narrower $x$-range relative to Fig.~1\,(a) to emphasize the decreasing range of electrostatic repulsion with increasing concentration; this is in agreement with Debye–Hückel theory.
\Cref{fig:Suppmat_Fig1_Alternative}\,(b) extends the $y$-range of Fig.~1\,(b) to highlight the weak electrostatic repulsion at the two lowest concentrations, \SI{0.001}{\Molar} and \SI{0.030}{\Molar}, revealing a clear re-entrant behavior of the electric double-layer force for \ch{LaCl3} concentrations above \SI{285}{m\Molar} (covered by the curve describing the \SI{0.177}{m\Molar} condition).

\begin{figure}[htb!]
    \begin{center}
    \includegraphics[width=2.53125in]{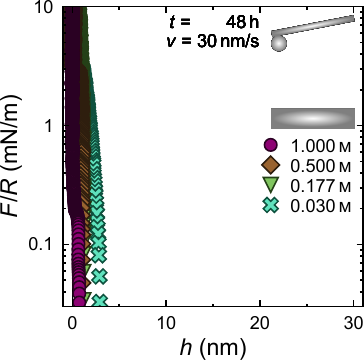}
    \caption{Normalized interaction force $F / R$ between the smaller silica colloidal probe with a radius of \SI{2.1}{\micro\meter} and a planar quartz substrate at separation $h$ immersed in aqueous \ch{LaCl3} electrolytes of various concentrations: \SIlist{0.030; 0.177; 0.500; 1.000}{\Molar}.
    The data were captured after an equilibration time of \SI{48}{\hour} with an approach speed of \SI{30}{\nano\metre/\second}.
    This figure is shown for direct comparison with Fig.~1\,(d) in the main text. \label{fig:Suppmat_SmallBeadFinal}}
    \end{center}
\end{figure}

One of the key outcomes from these investigations is the size of the colloidal probe, equilibration time, or approach speed alone do not control the emergence or magnitude of repulsive interactions at high \ch{LaCl3} concentrations. Rather, the synergy of these parameters governs the repulsive interactions. To illustrate the interplay of these parameters, \cref{fig:Suppmat_SmallBeadFinal} demonstrates the absence of anomalously long-ranged repulsive interactions for the small colloidal probe under the same conditions that produced clear anomalously long-ranged repulsion for the large probe in Fig.~1\,(d); i.e., long equilibration time, low approach velocities and \ch{LaCl3} concentrations above \SI{0.285}{M}.

\vfill

\clearpage

\section{Overlap of approach and retraction curves}

\begin{figure}[htb!]
    \begin{center}
    \includegraphics[width=5.0625in]{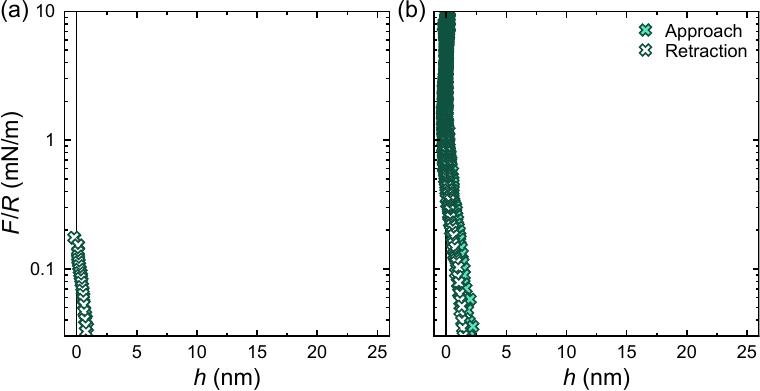}
    \caption{Normalized interaction force $F / R$ between the silica colloidal probe with \SI{8.5}{\micro\meter} radius and a planar quartz substrate at separation $h$ immersed in \SI{0.030}{\Molar} \ch{LaCl3}.
    The data were captured after an equilibration time of at least \SI{48}{\hour} with an approach speed of \SI{30}{\nano\metre/\second} at (a) a reduced setpoint of $\approx \SI{1.7}{\nano\newton}$ as well as (b) the standard setpoint of $\approx \SI{14}{\nano\newton}$. \label{fig:Suppmat_Approach_And_Retraction}}
    \end{center}
\end{figure}

Here we explore the impact of the setpoint on the measured force distance curves to verify that the measured interaction is reversible and free of loading-induced artifacts.
\Cref{fig:Suppmat_Approach_And_Retraction} presents the normalized interaction force $F/R$ between the \SI{8.5}{\micro\meter} silica colloidal probe and the planar quartz substrate in \SI{0.030}{\Molar} \ch{LaCl3}, recorded after an equilibration time of at least \SI{48}{\hour} at an approach speed of \SI{30}{\nano\metre\per\second}. 
Here both the approach and retraction curves are shown for two maximum setpoints: \Cref{fig:Suppmat_Approach_And_Retraction}\,(a) presents a reduced setpoint of $\approx \SI{1.7}{\nano\newton}$ and \Cref{fig:Suppmat_Approach_And_Retraction}\,(b) presents the standard setpoint of $\approx \SI{120}{\nano\newton}$.
In both cases, the approach and retraction traces overlap closely, with no measurable hysteresis, confirming that the long-ranged repulsion is an equilibrium feature of the electrolyte. 
The agreement between the two setpoints further demonstrates that the choice of setpoint does not influence the measured force profile.

\clearpage

\section{Modeling of the 1 millimolar dataset}

\begin{figure}[htb!]
    \centering
    \includegraphics[width=5.0625in]{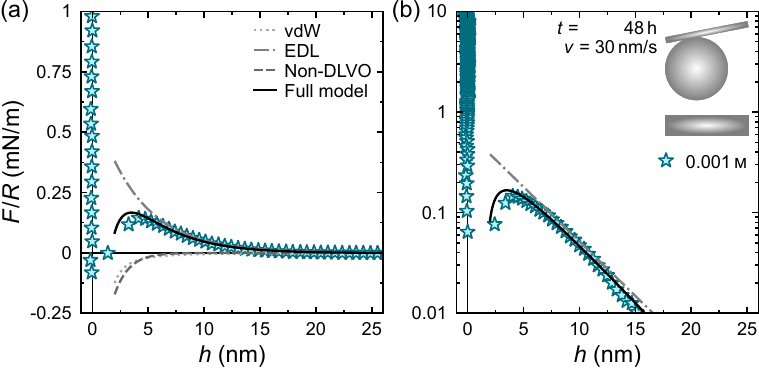}
    \caption{Normalized interaction force $F / R$ between a silica probe and a quartz substrate in \SI{0.001}{\Molar} \ch{LaCl3}. The same data as shown in Fig.~1\,(d) are presented together with the different contributions to the model of the interaction force. The DLVO contributions (van der Waals and electric double layer force) are complemented by an additional non-DLVO attraction \cite{Valmacco.2016b}.
    \label{fig:Suppmat_1mM}}
\end{figure}

As depicted in \Cref{fig:Suppmat_1mM}, an additional non-DLVO attraction $A_{ \mathrm{att} } \exp\left( h / \delta \right)$ needs to be taken into account for the modeling of the interaction force at the lowest \ch{LaCl3} concentration (\ie \SI{0.001}{\Molar}).
The amplitude \mbox{$A_{\mathrm{att}}\,=\,-\SI{0.8}{\milli\newton / \metre}$} and decay length $\delta = \SI{1.3}{\nano\metre}$ are chosen in accordance with the literature \cite{Valmacco.2016b}.
The origin of this attraction has been attributed to ion-ion correlation effects, possibly with additional contributions from surface charge heterogeneities and charge fluctuation forces \cite{Valmacco.2016b}.

\clearpage

\section{Scaling law}

\begin{figure}[htb!]
    \begin{center}
    \includegraphics[width=5.0625in]{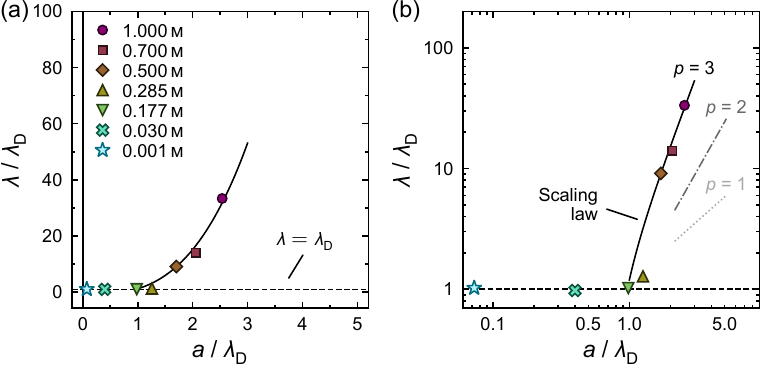}
    \caption{The experimental decay length $\lambda$, normalized by the Debye length $\lambda_{\mathrm{D}}$, plotted against the normalized mean ion size $a / \lambda_{\mathrm{D}}$ in (a) linear and (b) logarithmic representation (\cf Fig.~2 in the main text).
    The black solid line shows the scaling law from \cref{eqn:Suppmat_Scaling_law}, indicative for \emph{anomalous} underscreening, drawn for the parameters $A = 1.99$ and $B = -0.80$.
    The dark gray dash-dotted and light gray dotted lines denote the upper and lower bounds of the slope expected for \emph{normal} underscreening, respectively. \label{fig:Suppmat_Scaling_law}}
    \end{center}
\end{figure}

Motivated by the seminal analysis of Lee \emph{et al.} \cite{Lee.2017}, experimental decay lengths are often categorized according to the scaling law given in Eq.~(2) in the main text. Anomalous underscreening is characterized by a scaling exponent $p = 3$. Therefore, we have examined the validity of the relation
\begin{equation}
    \frac{\lambda}{\lambda_{\mathrm{D}}} = A \left(\frac{a}{\lambda_{\mathrm{D}}}\right)^3 + B \label{eqn:Suppmat_Scaling_law}
\end{equation}
for the data presented here. 
In line with previous investigations \cite{Robertson.2023, Lee.2017}, a mean ion size of $a = \SI{2.842}{\angstrom}$ was taken \cite{Shannon.1976}.
\Cref{fig:Suppmat_Scaling_law} shows the good agreement between \cref{eqn:Suppmat_Scaling_law} with $A = 1.99$ and $B = -0.80$ (solid black line) and the experimental data derived from the DLVO fits depicted in Fig.~1\,(d) for $a / \lambda_{\mathrm{D}} > 1$. 
It can be seen that the measured screening lengths are compatible with \emph{anomalous} underscreening, and clearly go beyond \emph{normal} underscreening, as is characterized by $p = 1$ (light gray dotted line) to $p = 2$ (dark gray dash-dotted line).

\clearpage

\section{Effective surface potentials}

\begin{figure}[htb!]
    \begin{center}
    \includegraphics[width=5.0625in]{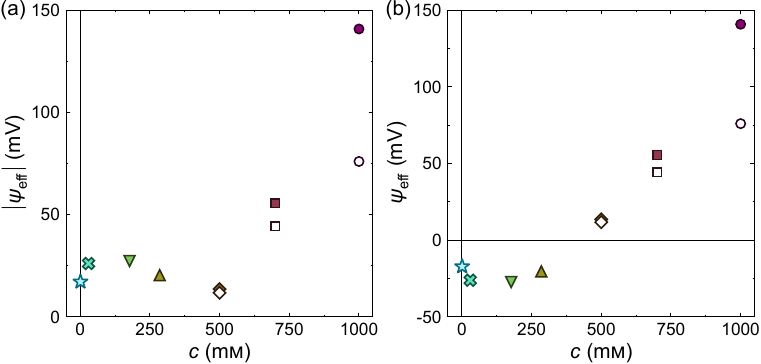}
    \caption{The effective surface potential $\psi_{\mathrm{eff}}$ as extracted from the DLVO fits depicted in Fig.~1\,(d) (closed symbols) as well as from the extended fits from Fig.~3 and \cref{fig:Suppmat_700mM} (open symbols) plotted against the electrolyte concentration. The symmetric model provides only the magnitude of the potential, not its sign. Therefore, (a) only shows the absolute value of $\psi_{\mathrm{eff}}$. However, we assume a charge reversal taking place between \qtylist{285; 500}{\milli\Molar} due to the adsorption of La$^{3+}$ ions at the negatively charged outer confining surfaces. This interpretation is depicted in (b).} \label{fig:Suppmat_Effective_Potential}
    \end{center}
\end{figure}

Including the electric double layer force (Eq.~(4)) in our modeling of the data allows to extract the effective potential $\psi_{\mathrm{eff}}$ of the surfaces. We note that $\psi_{\mathrm{eff}}$ characterizes the long-ranged decay of the electric double layer force and is quantitatively not identical to the potential at the origin of the diffuse layer. The symmetric model only allows to derive the absolute value of the potential (\cref{fig:Suppmat_Effective_Potential}\,(a)). Due to the strong adsorption of trivalent cations to negative interfaces, we interpret the drop in potential at intermediate concentrations as a charge reversal taking place between \qtylist{285; 500}{\milli\Molar} (\cref{fig:Suppmat_Effective_Potential}\,(b)) \cite{RuizCabello.2014}.

\clearpage

\section{Modeling of the oscillatory contribution to the interaction force}

As shown in Fig.~3, the agreement between model and experiment for \qtylist{0.500;1.000}{\Molar} \ch{LaCl3} is substantially improved by including a structural contribution to the interaction force. The complete model therefore consists of three components: the van der Waals force (${F_{\mathrm{vdW}}}/{R}$, Eq.~(3)) and electric double layer force (${F_{\mathrm{EDL}}}/{R}$, Eq.~(4)), which together constitute the DLVO interaction, and the structural force (${F_{\mathrm{struct}}}/{R}$, Eq.~(5)).

The fitting was carried out in two steps.
First, the DLVO contribution was determined by fitting the innermost region of the force curves, which appears linear in the semi-logarithmic representation of Fig.~3.
The Hamaker constant was to be in line with literature as $H = \SI{2.25(50)e-21}{\joule}$ \cite{Valmacco.2016}, while the effective surface potential $\psi_{\mathrm{eff}}$ and the decay length $\lambda$ were treated as fitting parameters.
In a second step, we followed the procedure proposed by Ludwig and von Klitzing \cite{Ludwig.2021b}.
The fitted DLVO contribution (blue dashed line in Fig.~3) was subtracted from the measured force curves, and the oscillatory component of the structural force was fitted to the resulting residual interaction.
The constant depletion attraction was then defined as the value of the oscillatory component at the transition distance $h^{\ast}$, yielding a continuous representation of the full structural force (red dotted line in Fig.~3).

\begin{figure}[htb!]
    \centering
    \includegraphics[width=2.53125in]{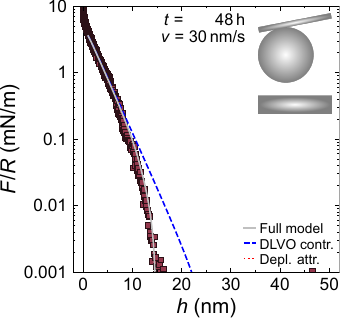}
    \caption{Normalized interaction force $F / R$ between a silica probe and a quartz substrate in \SI{0.700}{\Molar} \ch{LaCl3}. The same data as shown in Fig.~1\,(d) are presented over an extended force range, similar to Fig.~3 in the main text. An attractive depletion force (not visible), indicative of electrolyte structuring, is necessary to complement the DLVO contribution (dashed line). The full model (solid line) provides a proper description of the data throughout the entire force range. \label{fig:Suppmat_700mM}}
\end{figure}

The approach for \SI{0.700}{\Molar} \ch{LaCl3} was different.
While the data for \qtylist{0.500; 1.000}{\Molar} show a clear oscillation at low forces in Fig.~3, this feature is absent in \SI{0.700}{\Molar}.
We suggest that the missing oscillation in \SI{700}{mM} \ch{LaCl3} could be a result of the low degree of ordering between the ion clouds.
Therefore, only the depletion contribution of the structural force was modeled.
\cref{fig:Suppmat_700mM} demonstrates that the modeling of the data is significantly improved by the addition of this attractive depletion force, which clearly indicates the presence of a structure in the electrolyte near the interfaces despite the absence of oscillations.

\clearpage

\section{Relative permittivity of the electrolyte solutions}

\begin{figure}[htb!]
    \centering
    \includegraphics[width=2.53125in]{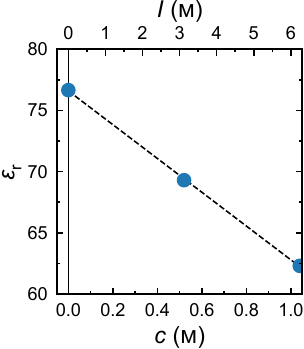}
    \caption{Relative permittivity $\varepsilon_{\mathrm{r}}$ of aqueous \ch{LaCl3} solutions at \SI{30}{\celsius} as a function of concentration $c$ and ionic strength $I$ (blue circles). 
    The permittivity for pure water at was averaged from ref. \cite{Wyman.1938, Malmberg.1956, Uematsu.1980}.
    The values for finite salt concentration were taken from ref. \cite{Hasted.1948} and corrected for temperature.
    The black dashed line indicates a linear model to the literature values.}
    \label{fig:Suppmat_Epsilon}
\end{figure}

In the concentration regime investigated (up to \SI{1}{\Molar}), the addition of \ch{LaCl3} leads to a decrease in the relative permittivity $\varepsilon_{\mathrm{r}}$ \cite{Hasted.1948}.
This reduction in $\varepsilon_{\mathrm{r}}$ directly enters the calculation of the Debye length in Eq.~(1) of the main text and must therefore be taken into account for a quantitative analysis of the screening length.
As can be seen in \cref{fig:Suppmat_Epsilon}, we estimated the concentration dependence of the relative permittivity via a linear interpolation to literature data for pure water and aqueous \ch{LaCl3} solutions of known concentration, yielding
\begin{equation}
    \varepsilon_{\mathrm{r}} \left( c \right) = 76.55 - 14.10 c. \label{eqn:Suppmat_Epsilon}
\end{equation}
To account for the elevated experimental temperature ($\SI{30}{\celsius}$ compared to $\SI{25}{\celsius}$), the literature values for non-vanishing \ch{LaCl3} concentrations were reduced by an offset of \num{1.7} units, in line with the known temperature-induced change for water.

\clearpage
\bibliography{bibliography}